\documentclass[latex,sn-mathphys]{sn-jnl}

\jyear{2022}%

\usepackage{lineno}

\begin{document}

\title[]{Nonlinear polarization holography of nanoscale iridium films}

\author[1]{\fnm{Mouli} \sur{Hazra}}
\author[2,3]{\fnm{Pallabi} \sur{Paul}}
\author[1]{\fnm{Doyeong} \sur{Kim}}
\author[4]{\fnm{Christin} \sur{David}}
\author[2,3,5]{\fnm{Stefanie} \sur{Gräfe}}
\author[4]{\fnm{Ulf} \sur{Peschel}}
\author[1]{\fnm{Matthias} \sur{Kübel}}
\author[2,3]{\fnm{Adriana} \sur{Szeghalmi}}
\author[1]{\fnm{Adrian N.} \sur{Pfeiffer}}

\affil[1]{\orgname{Friedrich Schiller University Jena}, \orgdiv{ Institute of Optics and Quantum Electronics, Abbe Center of Photonics}, \orgaddress{\street{Max-Wien-Platz 1}, \postcode{07743} \city{Jena}, \country{Germany}}}

\affil[2]{\orgname{Friedrich Schiller University Jena}, \orgdiv{Institute of Applied Physics, Abbe Center of Photonics}, \orgaddress{\street{Albert-Einstein-Straße 15}, \postcode{07745} \city{Jena}, \country{Germany}}}

\affil[3]{\orgname{Fraunhofer Institute for Applied Optics and Precision Engineering IOF}, \orgdiv{Center of Excellence in Photonics}, \orgaddress{\street{Albert-Einstein-Straße 7}, \postcode{07745} \city{Jena}, \country{Germany}}}

\affil[4]{\orgname{Friedrich Schiller University Jena}, \orgdiv{Institute of Condensed Matter Theory and Optics, Abbe Center of Photonics}, \orgaddress{\street{Max-Wien-Platz 1}, \postcode{07743} \city{Jena}, \country{Germany}}}

\affil[5]{\orgname{Friedrich Schiller University Jena}, \orgdiv{Institute of Physical Chemistry, Abbe Center of Photonics}, \orgaddress{\street{Max-Wien-Platz 1}, \postcode{07743} \city{Jena}, \country{Germany}}}





\maketitle

\section*{Abstract}

\textbf{
The phasing problem of heterodyne-detected nonlinear spectroscopy states that the relative time delay between the exciting pulses and a local oscillator must be known with subcycle precision to separate absorptive and dispersive contributions. Here, a solution to this problem is presented which is the time-domain analogue of holographic interferometry, in which the comparison of two holograms reveals changes of an objects size and position with interferometric precision (i.e. to fractions of a wavelength of light). The introduced method, called nonlinear polarization holography, provides equivalent information as attosecond nonlinear polarization spectroscopy but has the advantage of being all-optical instead of using an attosecond streak camera. Nonlinear polarization holography is used here to retrieve the time-domain nonlinear response of a nanoscale iridium film to an ultrashort femtosecond pulse. Using density matrix calculations it is shown that the knowledge of the nonlinear response with subcycle precision allows to distinguish excitation and relaxation mechanisms of low-energetic electrons that depend on the nanoscale structure of the iridium film. 
}

\section{Introduction}
The nonlinear polarization $P^{\mathrm{(NL)}}$ triggered by an intense laser pulse has been shown to reveal the dynamics of energy transfer for dielectrics using attosecond nonlinear polarization spectroscopy \cite{RN182, RN267}. In these measurements, $P^{\mathrm{(NL)}}$ is reconstructed with subcycle precision from the electric field of a laser pulse measured by an attosecond streak camera after propagation through the dielectric sample. All-optical methods of two-dimensional electronic spectroscopy \cite{RN286} could in principle deliver equivalent information, but in practice the so-called phasing problem complicates the retrieval of the nonlinear signal. The phasing problem (time-zero problem in the time-domain) arises because the relative time delay between the pulse sequence that triggers the nonlinear response and a reference pulse, called local oscillator, must be known with subcycle precision. Approaches to the phasing problem presented so far are based on calibrating the optical beam paths with interferometric precision \cite{RN287, RN288, RN289} or by post-processing the two-dimensional spectroscopy data using the projection-slice theorem when additionally pump-probe data is available \cite{RN291}. The disadvantages of these approaches include limited precision and a reduction of signal-to-noise ratio \cite{RN286}. Efforts have also been made to extend methods of laser pulse characterization to provide time-domain information about $P^{\mathrm{(NL)}}$. These methods usually assume a simple analytical relationship between the unknown laser pulse and the nonlinear response. For frequency-resolved optical gating, it was shown that an unknown parameter of the response function (the decay time) can be determined \cite{RN268}. 

For metals, measurements that assume the nonlinear polarization response to be instantaneous and proportional to the third power of the intensity have led to conflicting results in the past. The thus determined nonlinear refractive index $n_2$ has been shown to be strongly dependent on the pulse duration \cite{RN275}, indicating that the nonlinear response also has some delayed or cumulative features. This makes $P^{\mathrm{(NL)}}$ a promising observable to reveal ultrafast dynamics in metals; however, attosecond nonlinear polarization spectroscopy as for dielectrics has not yet been demonstrated for metals. Ultrafast dynamics in metals and nanoscale compounds of metals promise new applications in sensing, light harvesting and molecular interactions, especially if they can be controlled by light \cite{RN279}. Loss of coherence, caused for example by scattering, substantially affects electric excitations in metals, from direct currents to ultrafast light-matter interactions. The optical response in metals is strongly affected by scattering on impurities, surfaces and grain boundaries, which depends on the nanoscale structure \cite{RN277}. Electronic decoherence by collisions is evident in many material properties such as electrical and thermal resistivity \cite{RN276}. These processes involve low-energetic electrons in the conduction band, in contrast to the processes that have been studied by time-resolved two-photon photoemission experiments using femtosecond pulses \cite{RN274}, which scrutinize the relaxation mechanisms of optically excited (hot) electrons. Experiments using attosecond pulses for electron emission \cite{RN284, RN185, RN273, RN272} and transient absorption \cite{RN271}, on the other hand, give insight into the photoionization and screening dynamics. The properties of coherence loss of low-energetic electrons as deduced from electrical and thermal transport measurements are often summarized in a single quantity, namely the collision time \cite{RN274}. However, this is not sufficient for calculations of ultrafast light-matter interactions in metals, where the redistribution of a scattering event in k-space is crucial, as shown in this paper. 


\begin{figure}[ht]%
\centering
\includegraphics[width=0.7\textwidth]{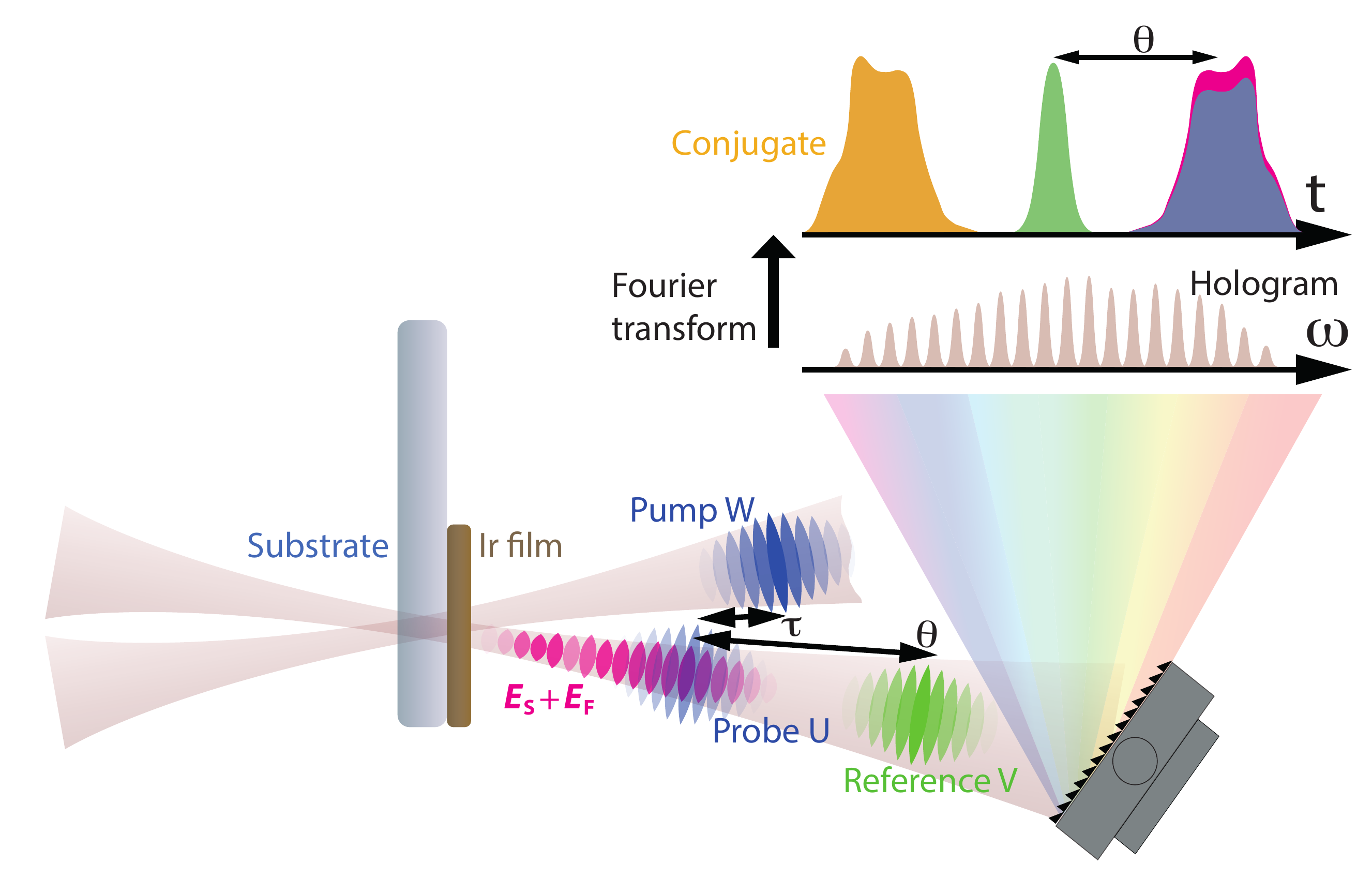}
\caption{\textbf{Nonlinear polarization holography.} Nonlinear polarization holography uses a probe wave U and a reference wave V with constant time delay $\theta$. The pump pulse W arrives at a variable delay $\tau$ at the sample, which is a dielectric substrate for the first hologram and additionally an Ir-film for the second hologram. The spectrally resolved intensities (the holograms) exhibit an interference pattern. $E_{\mathrm{S}}$ and $E_{\mathrm{F}}$, which are the fields emitted by the nonlinear responses of the substrate and the Ir-film, are contained in the second hologram, but only $E_{\mathrm{S}}$ is contained in the first hologram.}
\label{figHolo}
\end{figure}

Here, a variant of time-domain holography is introduced to determine the electric field emitted by $P^{\mathrm{(NL)}}$ along with the field of the laser pulse that triggers it. Time-domain holography has been used before to store digital data by using a reference beam and a data beam \cite{RN269}. The data beam can be realized by encoding digital data in a probe pulse by inducing a nonlinear response in a sample via cross-phase modulation (XPM) with a pump pulse. Unlike that, a near transform-limited pulse is used here as a pump to record two holograms, where the first sample is a substrate that is replaced by a substrate coated with an iridium (Ir) film for the second hologram. The goal is to determine the contribution of the Ir-film to the nonlinear signal. The method is the time-domain analogue of holographic interferometry, in which the comparison of two holograms reveals changes of an objects size and position with interferometric precision \cite{RN290}. The phasing problem of heterodyne-detected nonlinear spectroscopy is solved in this approach, because the time delay between the probe and the reference pulse (the local oscillator) is constant for both holograms and truncates in the retrieval. The approach is applied here to a pump-probe experiment and is suitable for any experimental implementation of heterodyne-detected nonlinear spectroscopy.

\section{Results}
Three near infrared (NIR) pulses interact in a sample with polarization perpendicular to the plane of incidence in pump-probe geometry (Fig.\,\ref{figHolo}). The pump pulse W with electric field $W(t)$ (peak intensity $I_{W}$ = 100\,GW/cm$^2$, beam waist 100\,$\mu$m) crosses a beam of two weak pulses (probe U and reference V with fields $U(t)$ and $V(t)$, peak intensity $I_{U} \approx I_{V} \approx$ 1\,GW/cm$^2$, beam waist 100\,$\mu$m) that have a delay of $\theta \approx 770$\,fs with a crossing angle $\alpha = 1.5^{\circ}$. U and V are produced by reflections of W from a 80-$\mu$m-thick fused silica plate at near-normal incidence, and $\theta$ equals the delay between front and rear reflections. The temporal shapes of the pulses are very similar (center wavelength: 800\,nm; pulse duration: 28\,fs), as they originate from the same laser pulse, but this is not a prerequisite for the functionality of the method; U and V only need to cover the same spectral range. 

For the first hologram, the sample is a 80-$\mu$m-thick substrate made from fused silica. Optical spectra are recorded for varying delay $\tau$ of W. This yields the hologram $I_{\mathrm{S}}(\omega,\tau) = \left\vert U(\omega) + V(\omega) + E_{\mathrm{S}}(\omega,\tau) \right\vert^2$, where $E_{\mathrm{S}}$ is the electric field emitted by the nonlinear polarization response of the substrate induced in course of the interaction with the pump W. The field emitted by the linear polarization is omitted in this expression for the sake of clearer equations; this field is independent of $\tau$ and therefore truncates in the holographic reconstruction (\ref{equ:S}, Appendix). 

For the second hologram, the sample is replaced by a substrate with a 5\,nm-thick Ir-film on the rear side produced by atomic layer deposition (ALD), see Appendix. To minimize the influence of thickness and repositioning aberrations, the same substrate is used for both measurements. The substrate is partially covered by the Ir-film and shifted perpendicular to the laser beams for the measurements with and without Ir-film (Fig.\,\ref{figHolo}). This yields the hologram $I_{\mathrm{SF}}(\omega,\tau) = \left\vert U(\omega) + V(\omega) + E_{\mathrm{S}}(\omega,\tau) +  E_{\mathrm{F}}(\omega,\tau) \right\vert^2 T(\omega)$. In this expression, the linear polarization of the Ir-film is accounted for by its linear intensity transmittance $T(\omega) = \left\vert t(\omega) \right\vert^2$ with the linear amplitude transmittance $t(\omega)$. The originally measured field emitted due to the nonlinear response of the Ir-film incident on the spectrometer is $E_{\mathrm{F}}(\omega) t(\omega)$, but, to allow for a comparison on equal footing we generate $E_{\mathrm{F}}$ by linear back-propagation through the Ir-film. 

The holographic reconstruction is a digital procedure which is analogous to the irradiation of a spatial hologram with the reference beam (see Appendix). The wave U is uniquely determined by the pulse retrieval using the first hologram except for the absolute phase and time zero \cite{RN192}, which carries on to the reconstruction of $E_{\mathrm{S}}$ and $E_{\mathrm{F}}$. The relative phases and times of $U(t)$, $E_{\mathrm{S}}(t)$ and $E_{\mathrm{F}}(t)$ are retrieved from the holograms. This is equivalent to the information provided by attosecond nonlinear polarization spectroscopy using an attosecond streak camera \cite{RN182}, except that the absolute phase is not determined in the present experiment. However, the absolute phase could be determined additionally if the method is used with stabilized carrier-envelope phase. For valid results of the pulse retrieval, the intensity must be sufficiently low to ensure that the interaction in the substrate is in the perturbative regime with instantaneous nonlinear response - a condition that is excellently met in the present case \cite{RN182}. However, the method makes no such assumption about the interaction in the metal film, meaning that also non-perturbative responses of the metal film can be retrieved. 

\begin{figure}[ht]%
\centering
\includegraphics[width=1\textwidth]{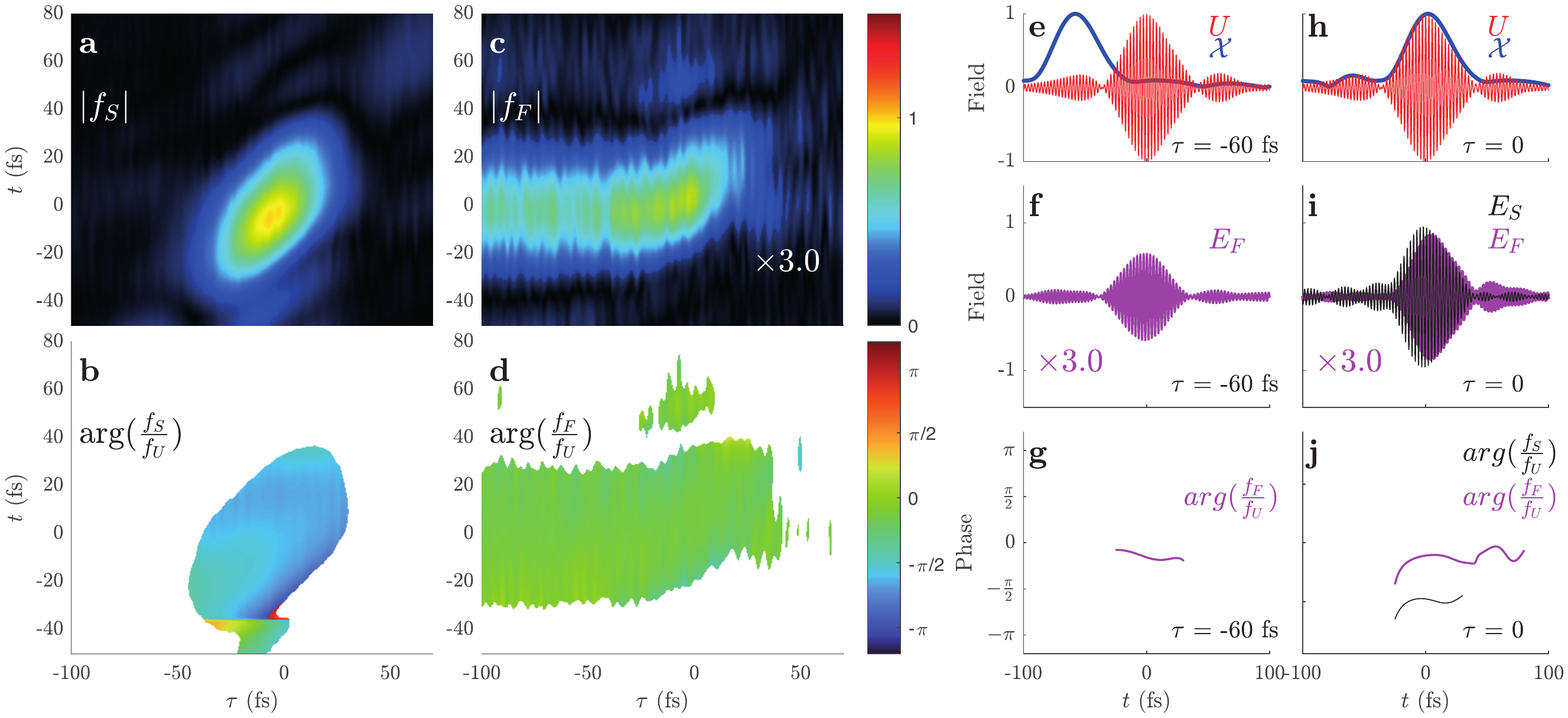}
\caption{\textbf{Nonlinear polarization response.} 
The envelope of the wave emitted by the nonlinear polarization response of the substrate is displayed in \textbf{a} and \textbf{b}. \textbf{c} and \textbf{d} show the corresponding quantities of the Ir-film. The amplitudes are displayed in \textbf{a} and \textbf{c}, the phase-lags with respect to the probe field $U$ in \textbf{b} and \textbf{d}. Curves are displayed for $\tau = -60$\,fs (\textbf{e}, \textbf{f}, \textbf{g}, the pump arrives prior to the probe pulse) and for $\tau = 0$ (\textbf{h}, \textbf{i}, \textbf{j}). \textbf{e} and \textbf{h} show the probe $U(t)$ (red) and the pump envelope $\mathcal{X}(t)$ (blue, see Appendix). \textbf{f} and \textbf{i} show the electric fields emitted due to the nonlinear responses; \textbf{g} and \textbf{j} show their phase-lags with respect to the probe pulse for the substrate (black) and for the Ir-film (magenta). $E_{\mathrm{S}}$ is very weak for $\tau = -60$\,fs and hence not displayed in \textbf{f} and \textbf{g}. In all panels, $E_{\mathrm{S}}$ and $E_{\mathrm{F}}$ have been normalized to the maximum of $E_{\mathrm{S}}$. For better comparison, $E_{\mathrm{F}}$ is scaled by 3, because $E_{\mathrm{S}}$ is about three times stronger than $E_{\mathrm{F}}$. The envelopes $f_{\mathrm{S}}$, $f_{\mathrm{F}}$ $f_{\mathrm{U}}$ are calculated from the fields according to $E_\mathtt{n}(t)= \Re \{ f_\mathtt{n}(t)\exp(i(\omega_0 t)) \}$ with $\omega_0 = 2.36$\,PHz and $n  = \mathrm{S},\mathrm{F},\mathrm{U}$. Only regions where the amplitude is greater than 15\% of the maximum are shown in \textbf{b}, \textbf{d}, \textbf{g}, \textbf{j}. 
}
\label{figData}
\end{figure}

The retrieved nonlinear response, presented in Fig.\,\ref{figData}, shows that a wave is emitted from the nonlinear polarization of the substrate only when U and W overlap temporally, see Fig.\,\ref{figData} \textbf{a} and \textbf{b}. The waveforms of $E_{\mathrm{S}}$ and $U$ are approximately the same, but $E_{\mathrm{S}}$ is temporally shifted with respect to U, always towards the pump pulse resulting in a tilted ellipse in Fig.\,\ref{figData}\,\textbf{a}. The phase is shifted with respect to $U$ by $-\pi/2$, i.e. the phase-lag between $E_{\mathrm{S}}$ and $U$ is a quarter of an optical cycle. Both, the pump induced delay as well as the phase shift of $E_{\mathrm{S}}$ are consistent with the action of XPM induced by an instantaneous nonlinearity. For monochromatic waves, this is commonly expressed by an intensity-dependent refractive index $n = n_0 + n_2 I$, known as the optical Kerr effect. With macroscopic propagation, focusing (self-focusing due to a Kerr lens) occurs. For a peak intensity of the pump-pulse of $I_{W}$ = 100\,GW/cm$^2$ this agrees with the literature value for $n_2$ of fused silica ($2.73 \times 10^{-20}$\,m$^2$/W \cite{RN235}). As our set-up generates reliable results for the well-known nonlinear response of fused silica we can now go on and measure the much less investigated nonlinear response of the Ir-film. 

The wave $E_{\mathrm{F}}$ is emitted by the nonlinear polarization of the Ir-film when the pump W arrives simultaneously with the probe U or earlier (negative delay), indicating a cumulative nonlinearity. For $\tau < -30$\,fs (pump before probe), the waveforms of $E_{\mathrm{F}}$ and $U$ are about the same. This is consistent with the assumption that the earlier-arriving pump changed the properties of the Ir-film, altering its complex refractive index. The phase-lag with respect to U is less than $\pi/2$, effectively amplifying U. In other words, the preceding pump saturates absorption and increases the transmission through the Ir-film. This bleaching does not decay within our measurement range of 100\,fs. For temporal overlap of probe U and pump W, further changes occur. We again see a temporal shift of $E_{\mathrm{F}}$ with respect to U indicating a partially instantaneous response similar as for the substrate. But, overall the waveforms of $E_{\mathrm{F}}$ and $U$ are significantly different. The pre-pulse of U is almost not visible in $E_{\mathrm{F}}$, whereas the post-pulse is considerably amplified due to the reduced absorption (Fig.\,\ref{figData}\textbf{i}). This results in an island that appears in Fig.\,\ref{figData}\,\textbf{c} at $\tau \approx 0$ and $t \approx 50$\,fs. This again indicates a delayed response, but now with a delay or relaxation time within the duration of the pulse. To sum up we have experimentally found strong hints for a mixture of instantaneous and delayed response in Ir-films. We will now shed light on the internal dynamics of these processes through numerical simulations, with a particular focus on fast relaxation phenomena.

\section{Discussion}
Time-dependent density-functional theory is often used to reproduce experimental data but is not suited to test different models of relaxation. Therefore, we compute the $k$-resolved density matrix (\cite{RN249}, see Appendix). All relaxation models discussed here produce an identical linear response and therefore cannot be distinguished by standard linear measurements of resistivity or refractive index. The main purpose of our calculations is to elucidate how the subcycle timing of the nonlinear response is affected by different relaxation mechanisms.

\begin{figure}[ht]%
\centering
\includegraphics[width=1\textwidth]{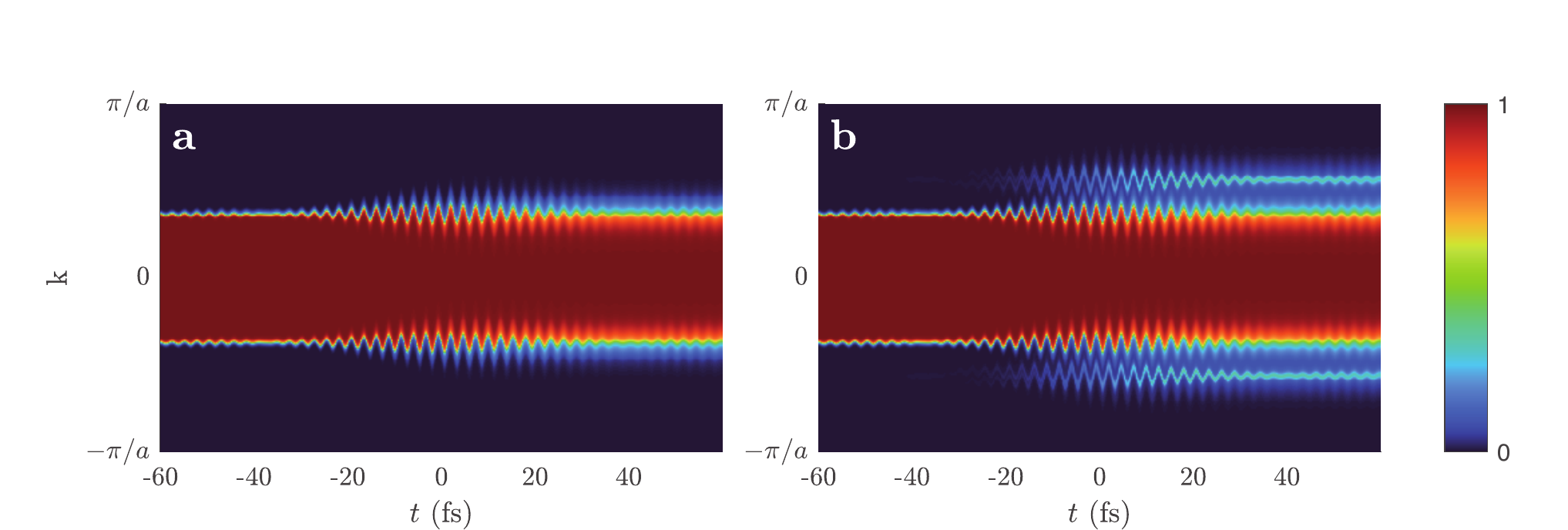}
\caption{\textbf{Cumulative mechanisms of nonlinearity.} \textbf{a} The distribution in $k$-space of the conduction band with symmetrizing decay (according to \ref{eq:mirror}, Appendix). \textbf{b} The distribution in $k$-space of the conduction band with symmetrizing decay and population transfer from the valence to the conduction band. 
}
\label{figMechanisms}
\end{figure}

Light-matter interaction in the Ir-film is dominated by two contributions: free carrier response in a half-filled conduction band (intraband transitions) and excitations of electrons from filled valence bands into a conduction band (interband transitions). The intraband transitions include an instantaneous (non-cumulative) nonlinearity, which is caused by the anharmonicity of the bands. The contribution of this non-cumulative nonlinearity is revealed by the increase of the amplitude of $E_{\mathrm{F}}$ in the region of temporal pulse overlap. The intraband transitions also include a cumulative nonlinearity, which is caused by the redistribution in $k$-space due to scattering of accelerated electrons towards a non-thermal electron distribution that gradually builds up during the laser pulse (Fig.\,\ref{figMechanisms}\,\textbf{a}). The width of this distribution in $k$-space is given by the excursion of the electrons, which is proportional to the laser electric field. Scattering on ions, impurities, grain boundaries and surfaces as it is is particularly important in ALD layers with their island-like structure is expected to be nearly elastic \cite{RN277, RN283}. Therefore, we implement this redistribution in $k$-space as a relaxation term that attenuates the current by symmetrizing the $k$-distribution but conserves the kinetic electron energy (see Appendix \ref{eq:mirror} and Ref.\,\cite{RN224}). 

\begin{figure}[ht]%
\centering
\includegraphics[width=1\textwidth]{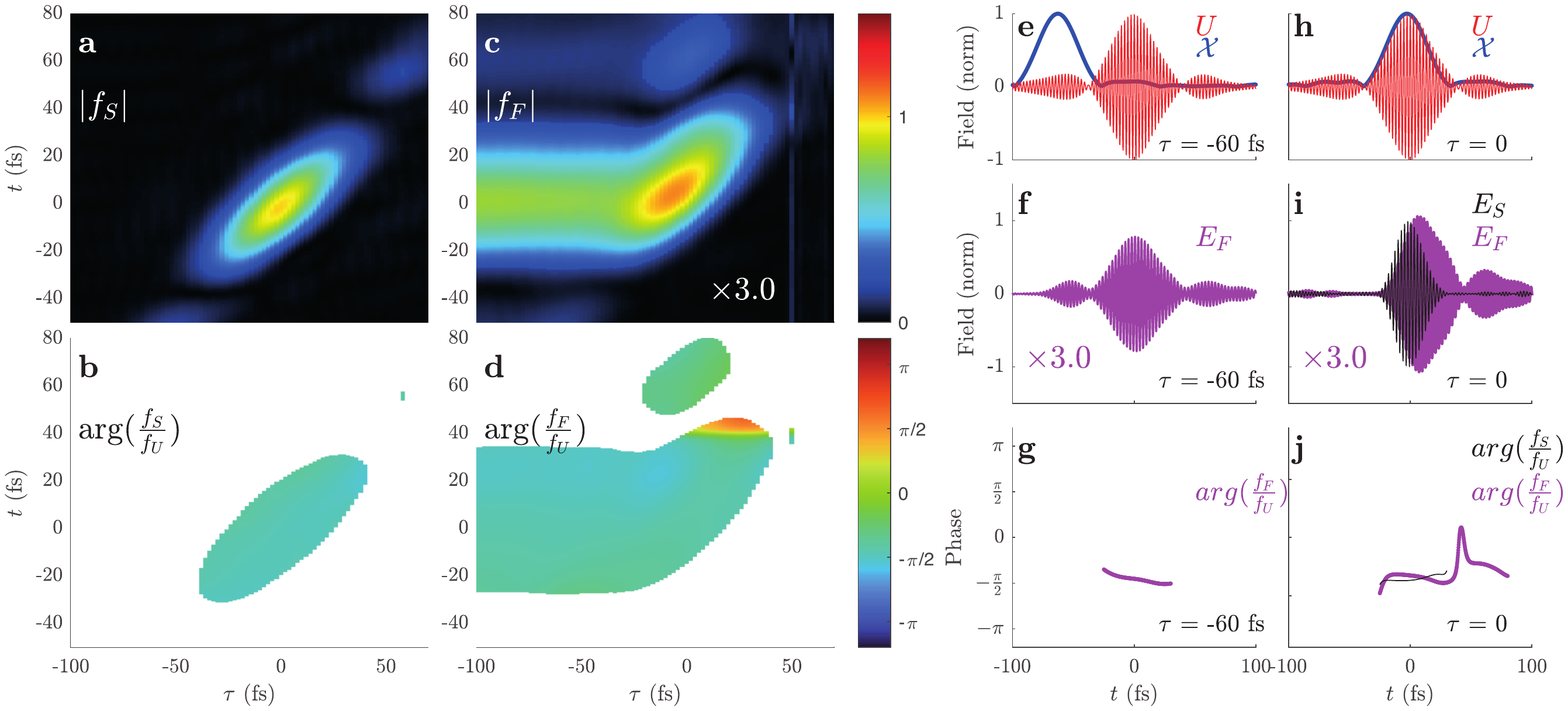}
\caption{\textbf{Simulation of the nonlinear polarization response.} The same as Fig.\,\ref{figData}, except that synthetic data is displayed using 2 bands, the symmetrizing decay (according to (\ref{eq:mirror})) and constant relaxation times. }
\label{figSim}
\end{figure}

A further cumulative nonlinearity is caused by interband transitions \cite{RN281}. During the pulse electrons gradually appear in the conduction band at $k$-positions where the energy gap is in resonance with the laser field (Fig.\,\ref{figMechanisms}\,\textbf{b} and Fig.\,\ref{figRefIdx}\,\textbf{a}). The calculation including one valence band and one conduction band and using the symmetrizing decay in $k$-space reproduces the data reasonably well (Fig.\,\ref{figSim}). For example, the increase of the amplitude of $E_{\mathrm{F}}$ in the region of temporal pulse overlap, attributed to intraband motion in anharmonic bands, is qualitatively reproduced. Also, the accumulation of the nonlinearity during the laser pulse, most evident in the form of an island appearing at $\tau \approx 0$ and $t \approx 50$\,fs, is consistent with the data (panel \textbf{c}). Even the $t$-$\tau$-dependent phase-lag shown in panel \textbf{d} agrees qualitatively within the scanned range of pump-probe delays, only the absolute value of the phase-lag is closer to $-\pi/2$ in the calculation than in the experimental data (panel \textbf{j}). 

There are further mechanisms of relaxation that are hitherto not included in our calculation. If only the symmetrizing redistribution in k-space is considered, neither the internal thermalization of the electron gas nor the equilibration of the electronic and lattice temperature (external thermalization) are reached. However, the internal thermalization is expected to be rather slow and comparable with the external thermalization for electrons in the vicinity of the Fermi level \cite{RN280, RN282}, in contrast to the fast decay of optically excited electrons due to electron-electron scattering \cite{RN274}. The comparison of the experimental data with a calculation in which the relaxation term is implemented as a relaxation into the equilibrium distribution (external thermalization) shows that this decay, which does not conserve the kinetic electron energy of the electron gas, is not observed within the time range of this study (see Appendix).

\section{Summary and outlook}
Each of the individual mechanisms that cause a nonlinear response in our model (see Appendix, Figs.\,\ref{fig1Band} and Fig.\,\ref{fig2Bands}) affects the subcycle timing of the nonlinear polarization in a characteristic way. Electron motion in anharmonic bands causes a phase-lag of $-\pi/2$, similar as observed in the fused silica substrate (focusing nonlinearity). Excitation of electrons from valence into conduction bands (interband nonlinearity) causes a phase-lag near $\pi/2$ (defocusing nonlinearity). Contrary to that, the symmetrizing decay (intraband nonlinearity) causes a phase-lag near $-\pi/2$. The contribution of intraband transitions to the linear refractive index (\ref{equ:intraS}, Appendix) suggests that the strength of the intraband nonlinearity depends on the electron group velocity at the Fermi edge and that the phase will deviate from $-\pi/2$ when the collision time $T_c$ is very short (comparable to the laser optical cycle). Similarly, the contribution of interband transitions to the refractive index (\ref{equ:interS}, Appendix) suggests that the strength of the interband nonlinearity depends on the dipole matrix element and that the phase will deviate from $\pi/2$ when the dephasing time $T_2$ is comparable to the laser optical cycle or the inverse transition frequency. Hence, the $t$-$\tau$-dependent plot of the phase-lag can be regarded as a fingerprint of the relaxation mechanism. 

In summary, nonlinear polarization holography is suited to retrieve the nonlinear response of a nanoscale metal film to an ultrashort laser pulse with femtosecond resolution. Instantaneous and cumulative contributions can be distinguished, and the gradual buildup of the nonlinear response during the laser pulse is observed. The subcycle timing of the nonlinear response with respect to the excitation pulses is particularly valuable for gaining insight into relaxation mechanisms. Different models of relaxation can be tested in calculations of the $k$-resolved density matrix. For the case of a nanoscale Ir-film, the nonlinear response on the time-scale of 100\,fs can be modeled surprisingly well using a rather simple model including a half-filled conduction and a valence band. A symmetrizing $k$-space relaxation, caused by prevailing
near-elastic scattering, proofs to influence the ultrafast nonlinear response significantly. A certain disagreement between the phases of experimentally observed and simulated fields hints to a more complex nature of scattering events, which will be investigated in further experiments.


\section{Appendix}

\subsection{Pulse retrieval}

The temporal shapes of the laser pulses are characterized using XPM scans, which is a recently demonstrated method for the simultaneous characterization of two unknown and independent laser pulses \cite{RN192}. The pulse retrieval is analytic, and the fidelity can be checked by comparing the complex-valued data trace with the retrieved trace. Originally demonstrated for the retrieval of two pulses in the deep ultraviolet, which are scanned with a further unknown NIR pulse \cite{RN194}, the method is here applied for three unknown frequency-degenerate pulses U, V and W using the variant \emph{center} of the original publication \cite{RN192}. 

The basis for the pulse retrieval is the complex trace $J_{\mathrm{S}}(\omega,\tau) - J_{\mathrm{S}}(\omega,\tau = \infty)$ (see Holographic reconstruction), followed by Fourier transform from $\tau$ to $\omega_{\tau}$. This yields the trace $\textbf{Y}(\omega,\omega_{\tau})$, which 
\begin{equation} 
\textbf{Y}(\omega,\omega_{\tau}) \propto U(\omega + \omega_{\tau})V^*(\omega)\mathcal{X}(-\omega_{\tau}).
\label{eq:Y9}
\end{equation}
with $\mathcal{X}(\omega_{\tau}) = \int d\omega_2R(\omega_2)R(\omega_{\tau}-\omega_2)$. This data traces is of the advantageous kind that it factorizes in three functions that only depend on $\omega$, $\omega_{\tau}$ and ${\omega + \omega_{\tau}}$ respectively. For the numerical retrieval of $U$ and $V$, it is advantageous to perform the substitution $\omega_{\tau}$ $\rightarrow$ $-\omega - \omega_{\tau}$ and to remove the fast phase oscillations of $\textbf{Y}_{\mathrm{e}}$ by multiplication with $\mathrm{e}^{-i \omega_{\tau} \theta}$. This results in
\begin{equation}
\bar{\textbf{Y}}(\omega, \omega_{\tau}) \propto U(-\omega_{\tau})V^*(\omega)\mathcal{X}(\omega_{\tau} + \omega).
\label{eq:Z9b}
\end{equation}
As detailed in Refs. \cite{RN31, RN184}, preliminary solutions $\hat{U}(\omega)$ and $\hat{V}(\omega)$ are obtained by logarithmic differentiation with respect to $\omega$ respectively $\omega_{\tau}$ and subsequent exponential integration on the diagonal of the data trace. These are connected to the true solutions by $U(\omega) = \hat{U}(\omega)\mathrm{e}^{i \alpha + i T \omega + s \omega}$ and $V(\omega) = \hat{V}(\omega)\mathrm{e}^{i \beta + i T \omega + s \omega}$ with the real integration constants $\alpha, \beta, T$ and $s$. The absolute arrival time $T$ cannot be determined from the data traces, and also the determination of both carrier envelope phases $\alpha$ and $\beta$ individually is impossible. These parameters are sometimes referred to as \textit{trivial ambiguities}. The determination of $s$, which affects the spectra $\left\vert U(\omega) \right\vert^2$ and $\left\vert V(\omega) \right\vert^2$, is necessary for pulse retrieval. In the present case, it is straightforward to determine $s$ because the spectra are known and $\left\vert U(\omega) \right\vert^2 = \left\vert V(\omega) \right\vert^2$. 

\begin{figure}[htbp]%
\centering
\includegraphics[width=1\textwidth]{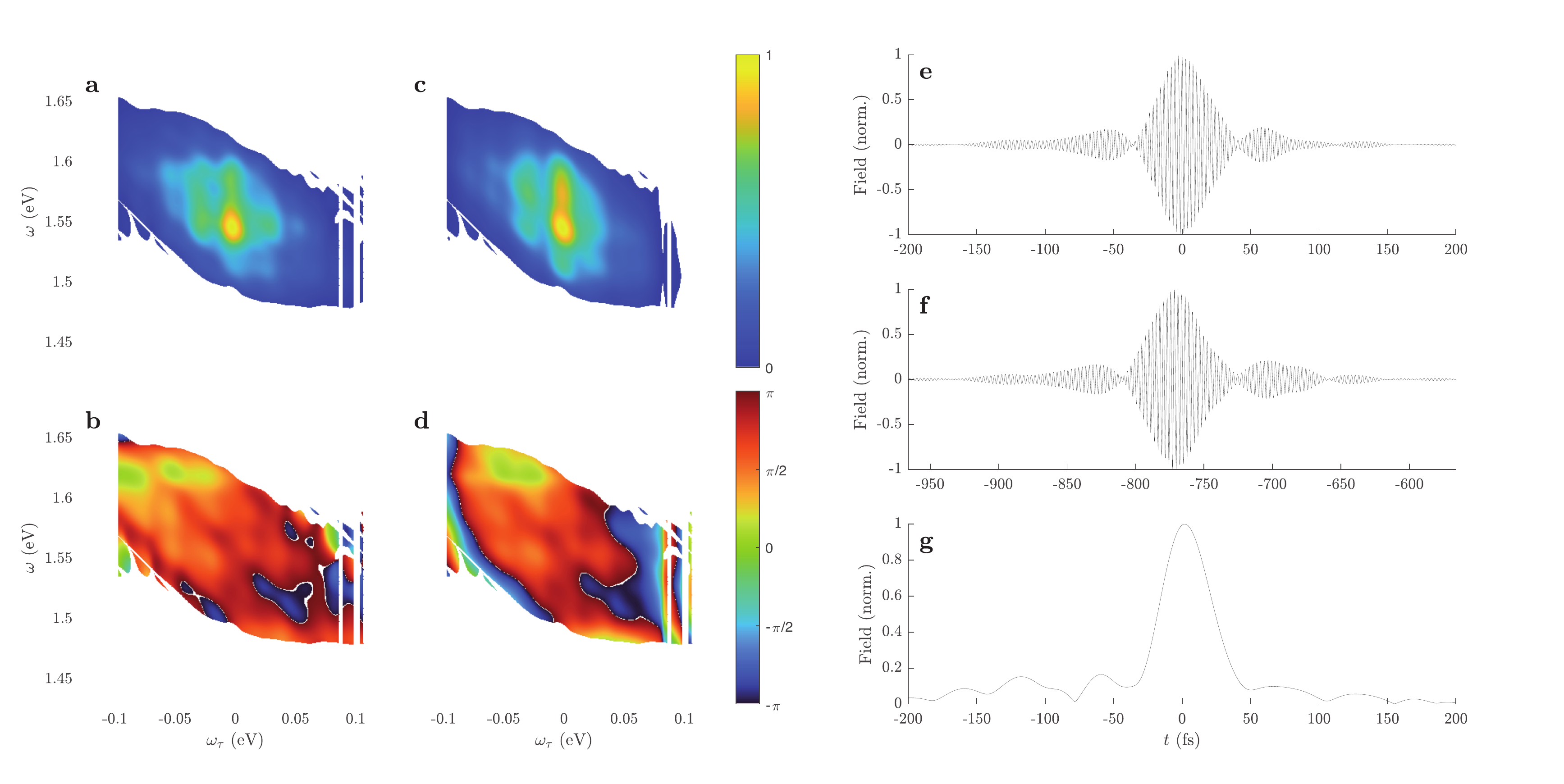}
\caption{\textbf{Pulse retrieval}. amplitudes (\textbf{a} and \textbf{c}) and phases (\textbf{b} and \textbf{d}) of $\textbf{Y}(\omega,\omega_\tau)$. The measured trace is depicted in \textbf{a} and \textbf{c}, the retrieved trace in \textbf{c} and \textbf{d}. The traces are only shown in the region where the amplitude exceeds 10\% of the maximum.
The retrieved electric fields of $U(t)$, $V(t)$ and $\mathcal{X}(t)$ are displayed in \textbf{e}, \textbf{f} and \textbf{g}.
}
\label{figPulseRetrieval}
\end{figure}

The pulse retrieval from $\textbf{Y}(\omega,\omega_\mathtt{R})$ is analytic, and the fidelity of the retrieval is checked by comparing the complex-valued data trace with the retrieved trace. The traces and the retrieved pulses are depicted in Fig.\,\ref{figPulseRetrieval}.

\subsection{Holographic reconstruction}

The holograms are inverse Fourier transformed, the side peak (alternating component) is cut-out and shifted to zero, and thereafter Fourier transformed.
The convention of the Fourier transform used here is  
\begin{equation}
\mathcal{F} \{ f(t) \} \propto \int_{-\infty}^{+\infty} f(t) \mathrm{e}^{-i \omega t} \mathrm{d}t.
\label{eq:FTconv}
\end{equation}
This digital procedure is analogous to the irradiation of a spatial hologram with the reference beam. The shift toward zero in pseudotime should be approximately $\theta$, but it is not necessary to know $\theta$ precisely, since it truncates in (\ref{equ:S}) and (\ref{equ:M}). This yields the complex-valued traces $J_{\mathrm{S}}(\omega,\tau) = \left( U(\omega) + E_{\mathrm{S}}(\omega,\tau)  \right) V^*(\omega) e^{-i \omega \theta}$ and 
$J_{\mathrm{SF}}(\omega,\tau) = \left( U(\omega) + E_{\mathrm{S}}(\omega,\tau) + E_{\mathrm{F}}(\omega,\tau) \right) V^*(\omega) e^{-i \omega \theta} T(\omega)$.
The temporal range of $\tau$ must only encompass the temporal range of $U$, $E_{\mathrm{S}}$ and $E_{\mathrm{F}}$, not the reference $V$. Using the complex-valued traces, the fields emitted by the nonlinear responses can be determined:
\begin{align}
E_{\mathrm{S}}(\omega,\tau) &= U(\omega) \left( \frac{J_{\mathrm{S}}(\omega,\tau)}{J_{\mathrm{S}}(\omega,\tau = \infty)} - 1 \right)
\label{equ:S} 
\\
E_{\mathrm{F}}(\omega,\tau) &= \left( U(\omega) + E_{\mathrm{S}}(\omega,\tau) \right) \left( \frac{J_{\mathrm{SF}}(\omega,\tau)}{J_{\mathrm{S}}(\omega,\tau) T(\omega)} - 1 \right)
. \label{equ:M} 
\end{align}
Here it is assumed that no nonlinear response is induced when the pump arrives after the probe ($E_{\mathrm{S}}(\omega,\tau = \infty) = E_{\mathrm{F}}(\omega,\tau = \infty) = 0$). In the analysis, the averaged data in the range $\tau > 70$\,fs is used for $J_{\mathrm{S}}(\omega,\tau = \infty)$ and $J_{\mathrm{F}}(\omega,\tau = \infty)$. The transmission is determined by $T(\omega) = \frac{J_{\mathrm{SF}}(\omega,\tau = \infty)}{J_{\mathrm{S}}(\omega,\tau = \infty)}$. The contribution of $E_{\mathrm{S}}$ in (\ref{equ:M}) can be neglected in the present case, because $E_{\mathrm{S}} \ll U$. 

The wave of U is uniquely determined by the pulse retrieval except for the absolute phase and time zero, which carries on to $E_{\mathrm{S}}$ and $E_{\mathrm{F}}$ in (\ref{equ:S}) and (\ref{equ:M}). However, the relative phases and times of $U(t)$, $E_{\mathrm{S}}(t)$ and $E_{\mathrm{F}}(t)$ are retrieved from the holograms, similar as the relative phases and times between $U(t)$, $V(t)$ and pump envelope $\mathcal{X}(t)$ are uniquely determined by the pulse retrieval. For valid results of the pulse retrieval, the intensity must be sufficiently low to ensure that the interaction in the substrate is in the perturbative regime with instantaneous nonlinear response, a condition that is excellently met in the present case \cite{RN182}. However, the method makes no such assumption about the interaction in the metal.

\subsection{Density matrix calculations}

\begin{figure}[htbp]%
\centering
\includegraphics[width=0.7\textwidth]{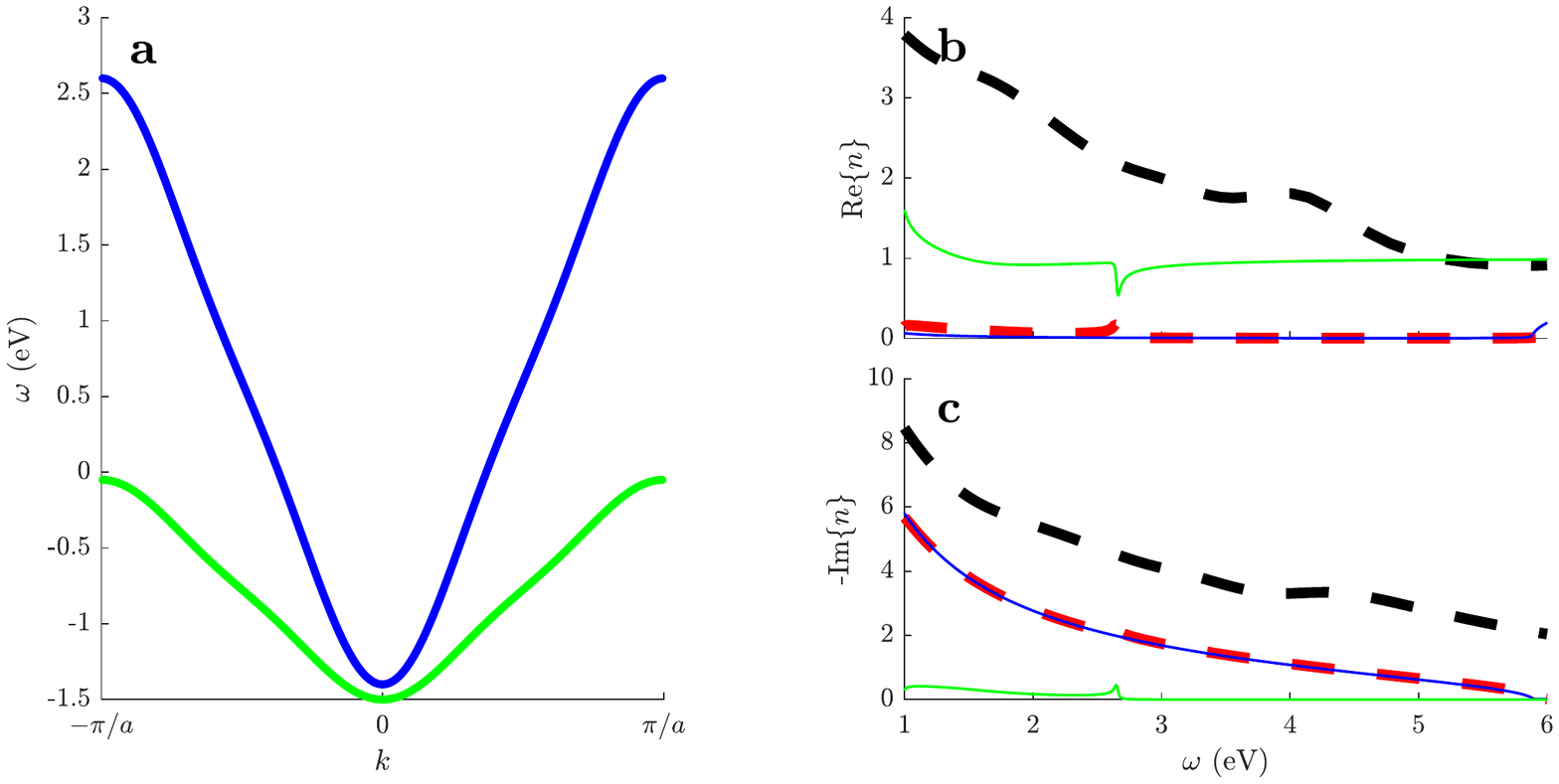}
\caption{\textbf{Band structure and refractive index}. \textbf{a} The calculations involve one conduction band (blue) and one valence band (green). The band shapes are approximated by $\omega^k = \omega_0 + \frac{1}{2} b_1 (1 - \mathrm{cos}(ka) - b_3 \left( \mathrm{cos}(3ka) - \mathrm{cos}(ka) \right) )$ with $\omega_0 = -1.4$\,eV, $b_1 = 4$\,eV, $b_3 = 2$ for the conduction band and $\omega_0 = -1.5$\,eV, $b_1 = 1.45$\,eV, $b_3 = 2$ for the valence band with the lattice constant $a = 0.384$\,nm. The shape of the conduction band is similar as calculations on https://materialsproject.org, the shape of the valence band is chosen to enable interband transitions within the optical spectrum of the laser pulse. The real (imaginary) part of the linear refractive index of the two-band model (\ref{equ:n}) (red) and experimental reference data \cite{RN278} (black dashed) are shown in \textbf{b} (\textbf{c}). The blue line shows the refractive index solely due to intra-band transitions in the conduction band (calculated with (\ref{equ:intraS})), the green line shows the refractive index solely due to inter-band transitions (calculated with (\ref{equ:interS})).
}
\label{figRefIdx}
\end{figure}

The evolution of the $k$-resolved density matrix is calculated by \cite{RN249, RN222}
\begin{align}
i \frac{d}{dt} \rho_{nm}^{k} = -\omega_{nm}^{k} \rho_{nm}^{k} 
+ E \cdot \sum_{l} \left( d_{lm}^{k} \rho_{nl}^{k} - d_{nl}^{k} \rho_{lm}^{k} \right) \nonumber\\
+ i E \cdot \frac{d}{dk} \rho_{nm}^{k}
+  i \left( \partial_t \rho_{nm}^{k}  \right)_{relax}
\label{LengthGauge} 
\end{align}
(atomic units are used). These equations are identical to the semiconductor Bloch equations, except that the Coulomb interaction is omitted and only one spatial dimension (the dimension of the polarization of the electric field) is considered \cite{RN218}. 
The diagonal elements $\rho_{nn}^k$ of the density matrix are the populations of the electronic bands, the off-diagonal elements $\rho_{nm}^k (n \neq m)$ are the coherences between the states with transition energies $\omega_{nm}^k = \omega_{m}^k - \omega_{n}^k$. The terms proportional to the dipole matrix elements $d_{nm}^{k}$ describe interband transitions. The terms proportional to $\frac{d}{dk}$ accelerate the electrons and holes within the bands which is called intraband transitions. To ease the calculations, the coordinate transform $k \to k + A$ is applied to (\ref{LengthGauge}), where $A$ is the vector potential defined by $E = - \partial_t A$:
\begin{align}
i \frac{d}{dt} \rho_{nm}^{k+A} = -\omega_{nm}^{k+A} \rho_{nm}^{k+A}
+ E \cdot \sum_{l} \left( d_{lm}^{k+A} \rho_{nl}^{k+A} - d_{nl}^{k+A} \rho_{lm}^{k+A} \right) \nonumber\\
+ i \left( \partial_t  \rho_{nm}^{k+A} \right)_{relax}
. \label{VelGauge} 
\end{align}
The initial band population at $t = - \infty$ is implemented as the Fermi–Dirac distribution at room temperature (293\,K). The calculations are based on one conduction band and optionally one valence band as defined in Fig.\,\ref{figRefIdx}. The valence to conduction band transitions $d_{12}^{k} = d_{21}^{k}$ are implemented as \cite{RN218}
\begin{equation}
d_{12}^k = d_{12}^{k=0} \frac{\omega_{12}^{k=0}}{\omega_{12}^k}
. \label{DipoleElement} 
\end{equation}
with $d_{12}^{k=0} = 10$\,au. 

Whereas the band structure parameters $\omega_{nm}^k$ and $d_{nm}^k$ can in principle accurately be calculated by \textit{ab initio} methods like density functional theory, the relaxation term $\left( \partial_t \rho_{nm}^k  \right)_{relax}$ is usually implemented as a phenomenological damping term and suited to test assumptions about the relaxation mechanism. Spontaneous decay of the conduction band electrons to the valence band is neglected here, because the lifetime in conduction bands (the $T_1$ time) usually exceeds 100\,fs \cite{RN281}. Collisions are included using the collision time $T_c$ by either
\begin{align}
\left( \partial_t \rho_{nn}^k  \right)_{relax}
= - \frac{1}{2 T_c} \left( \rho_{nn}^{k} - \rho_{nn}^{-k} \right)
 \label{eq:mirror} 
\end{align}
or
\begin{align}
\left( \partial_t \rho_{nn}^k  \right)_{relax}
= - \frac{1}{T_c} \left( \rho_{nn}^{k} - f_{\mathtt{FD}}(\omega_n^k)   \frac{\sum_k \rho_{nn}^{k}}{\sum_k f_{\mathtt{FD}}(\omega_n^k) }    \right)
. \label{eq:even} 
\end{align}
The approach in (\ref{eq:mirror}) follows Ref.\,\cite{RN224} and attenuates the current by symmetrizing the $\rho_{nn}^{k}$ distribution. 
Alternatively, the approach in (\ref{eq:even}) attenuates the current by decay directly into the equilibrium distribution, where $f_{\mathtt{FD}}$ is the Fermi–Dirac distribution at room temperature (293\,K). 

For the coherences,  
\begin{equation}
\left( \partial_t \rho_{nm}^k  \right)_{relax}
= - \frac{1}{T_2} \rho_{nm}^{k}
. \label{T2} 
\end{equation}
where $T_2$ is the interband dephasing time. Here it is assumed that $T_2$ is identical for all coherences and independent of $k$. The relation between the dephasing time $T_2$ and the collision time $T_c$ is not known. Here, $T_2 = 2 T_c$ is assumed, following the phenomenological picture that if scattering occurs to an electron at position $k$, its interband- and intraband coherences are likewise destroyed. 

The polarization $P$ and the current $J$ are calculated by 
\begin{align}
P &= N_e \sum_{n \ne m} \sum_k d_{nm} \rho_{nm}^k \delta k \label{equ:P} 
\\
J &= - N_e \sum_{n} \sum_k \rho_{nn}^k  v_{n}^k \delta k
, \label{equ:J} 
\end{align}
where $v_{n}^k = \partial_k \omega_{n}^k$ is the electron group velocity and $\delta k$ is the spacing in the $k$-grid. The macroscopic observables are scaled to an effective electron number density $N_e$, because the two-band approximation in one dimension does not reflect all contributing electrons. 

The resistivity of this model is
\begin{equation}
\rho
= \frac{1}{N_e T_c \left( v_{1}^{k_r} - v_{1}^{k_l} \right)}
, \label{equ:resist} 
\end{equation}
where $v_{1}^{k_r}$ and $v_{1}^{k_l}$ is the electron group velocity of the conduction band at the right and the left Fermi edge. For the present calculations, $N_e = 0.0045$ is chosen which reproduces the literature value $\rho = 47$\,n$\Omega$m\,$=0.22$\,au \cite{RN270} for $T_c = 30$\,fs, which is a typical value for bulk metals \cite{RN274}. For the Ir-film, the collision time $T_c = 15$\,fs is assumed. This value is lower than the value assumed for the bulk material to account for the nanoscale structure of the Ir-film \cite{RN277}. 

With the linear susceptibility of the intraband transitions
\begin{equation}
\chi^{(1)}_{\mathtt{intra}}(\omega)
= \frac{\mathcal{F} \left\{ J  \right\}}{i \omega  \mathcal{F} \left\{ E  \right\}}
= N_e \frac{v_{1}^{k_r} - v_{1}^{k_l}}{i \omega / T_c -\omega^2}
 \label{equ:intraS} 
\end{equation}
and the linear susceptibility of the interband transitions
\begin{equation}
\chi^{(1)}_{\mathtt{inter}}(\omega)
= \frac{\mathcal{F} \left\{ P  \right\}}{\mathcal{F} \left\{ E  \right\}}
= N_e \sum_k \frac{2 \omega_{12}^{k} ({d}_{12}^{k})^2 }{(\omega_{12}^{k})^2 -\omega^2 + 1/T_2^2 + 2 i \omega / T_2} \delta k
\label{equ:interS} 
\end{equation}
the linear refractive index is calculated as depicted in Fig.\,\ref{figRefIdx}:
\begin{equation}
n(\omega)
= \sqrt{1 + 4 \pi \chi^{(1)}_{\mathtt{inter}}(\omega) + 4 \pi \chi^{(1)}_{\mathtt{intra}}(\omega)}
. \label{equ:n} 
\end{equation}

The numerical calculations are performed on a $k$-grid with 37 points. The time-domain integration is performed using the 4th-order Runge–Kutta (RK4) method on a $t$-grid with 12001 points in the interval [-750, 750]\,fs.

\newpage
\begin{figure}[ht]%
\small 
Calculation 1: 1 band; equilibrium decay
\centering
\newline
\includegraphics[width=1\textwidth]{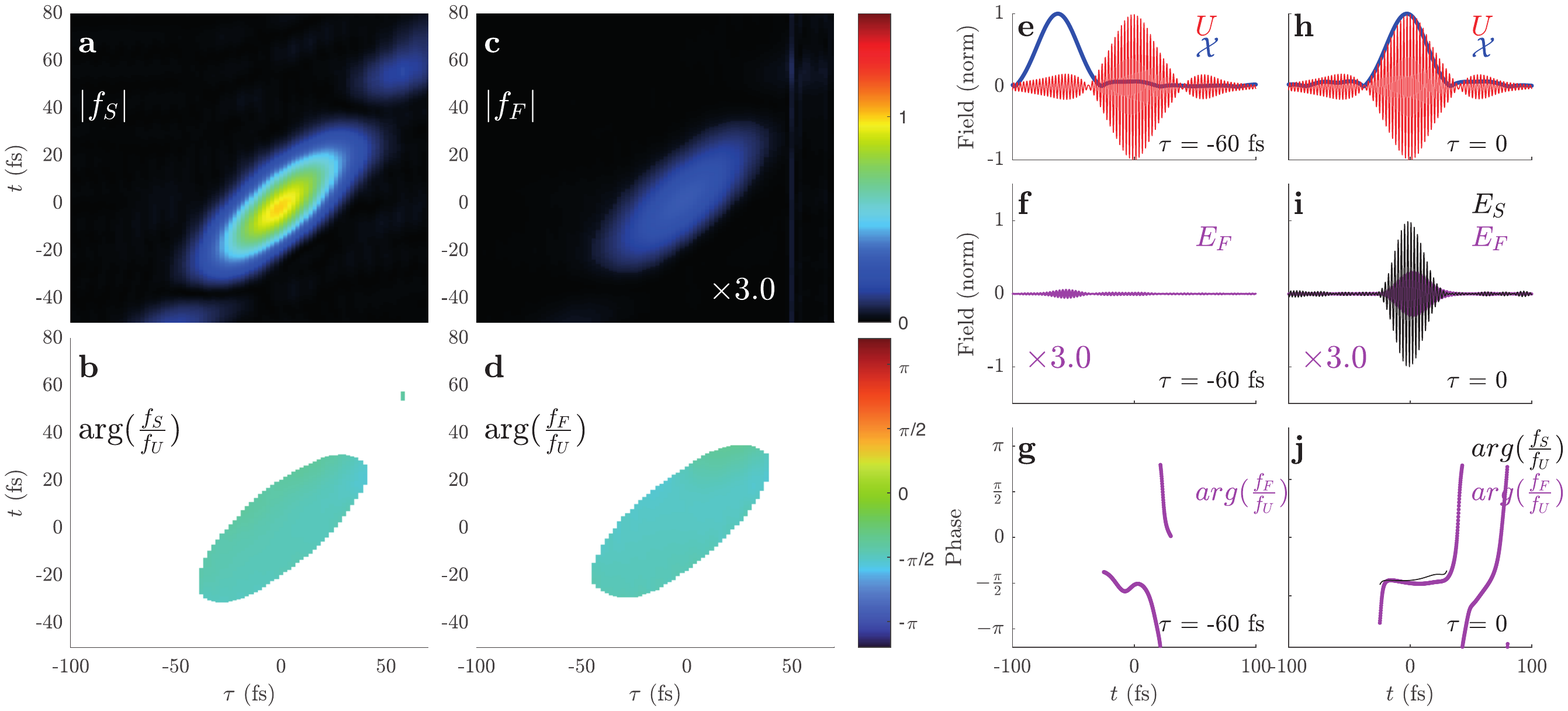}
\newline
Calculation 2: 1 band; symmetrizing decay
\centering
\newline
\includegraphics[width=1\textwidth]{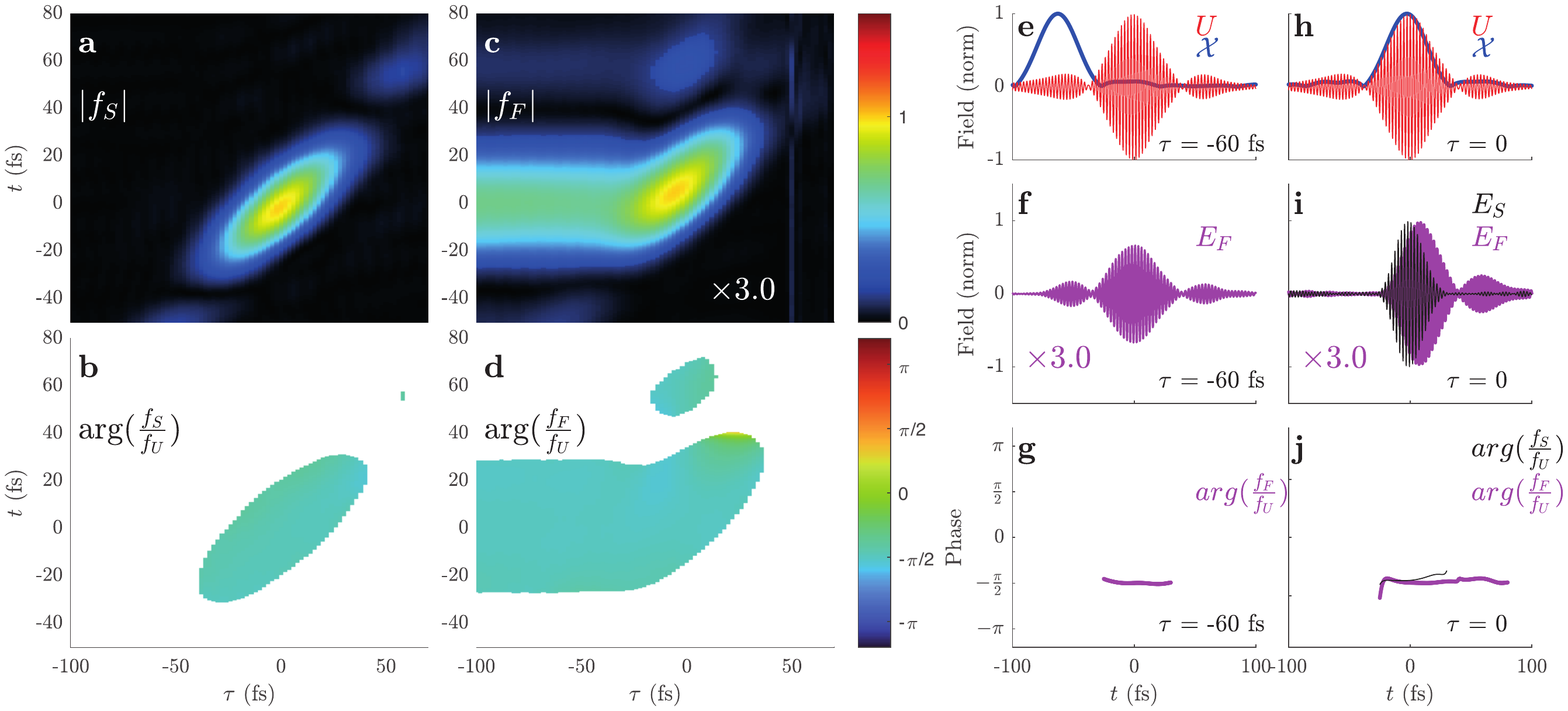}
\newline
\caption{\textbf{Comparison of calculations with 1 band}. Calculations including only the conduction band (see Fig.\,\ref{figRefIdx}) are displayed with equilibrium decay (according to (\ref{eq:even})) or symmetrizing decay (according to (\ref{eq:mirror})). For each calculation, \textbf{a} - \textbf{j} correspond to the panels in Fig.\,4 in the main paper.
}
\label{fig1Band}
\end{figure}

\newpage
\begin{figure}[ht]%
\small 
Calculation 1: 2 bands; equilibrium decay
\centering
\newline
\includegraphics[width=1\textwidth]{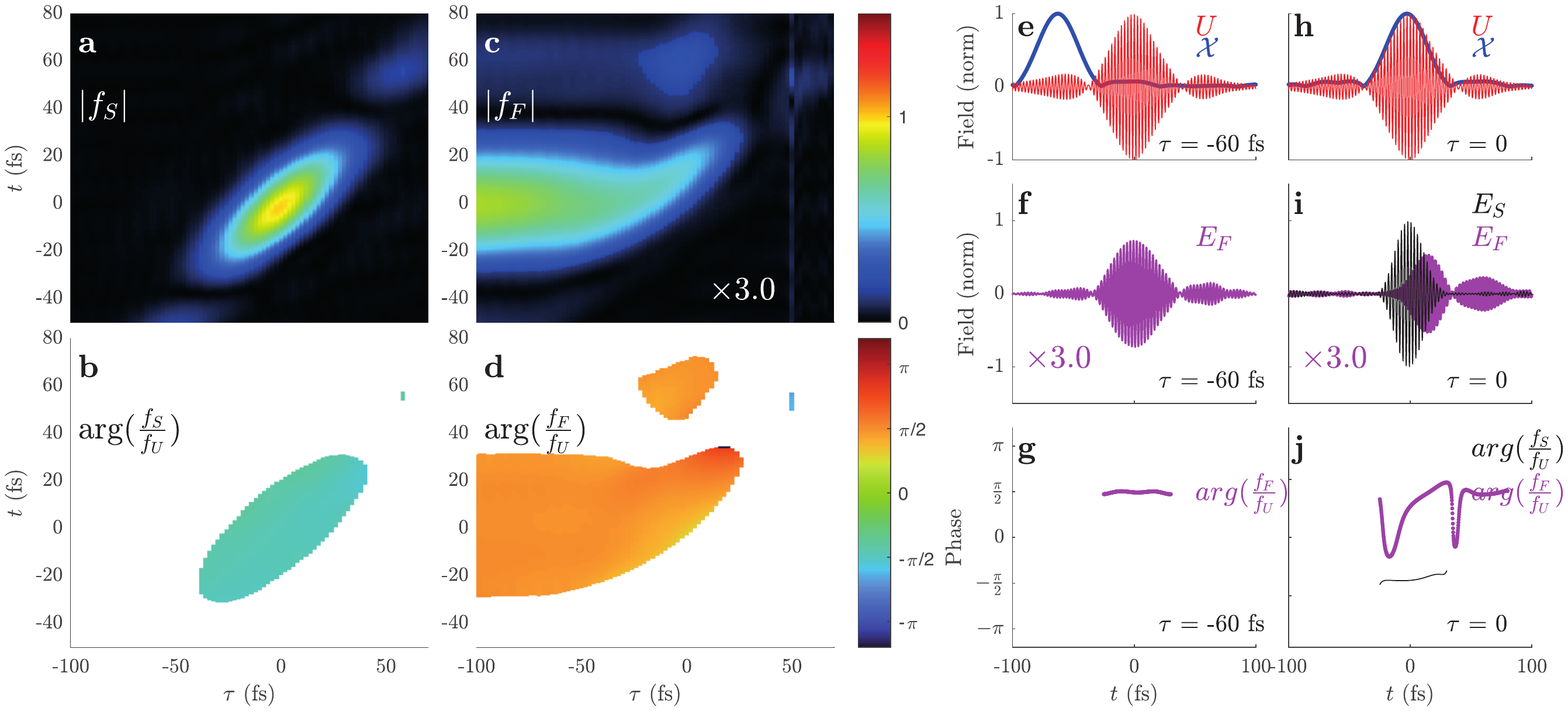}
\newline
Calculation 2: 2 bands; symmetrizing decay
\centering
\newline
\includegraphics[width=1\textwidth]{XPMS_Kerr_SBE_thick80_Int00.00_IntB00.10_FWHM25_GVD0_refidxn_SiO2_o_n_Iridium_2_mirror_Tdeph30.0_screennone.eps}
\newline
\caption{\textbf{Comparison of calculations with 2 bands}. The same as Fig.\,\ref{fig1Band}, except that also one valence band (see Fig.\,\ref{figRefIdx}) is included. Calculation 2 is identical to the calculation shown in the main text (Fig.\,4 in the main paper). 
}
\label{fig2Bands}
\end{figure}

\subsection{Macroscopic pulse propagation}

Macroscopic pulse propagation is calculated using the unidirectional pulse propagation equation (UPPE) \cite{RN141} through the entire sample, consisting of a substrate for the calculation of the first hologram and additionally an Ir-film for the calculation of the second hologram. For the substrate, instantaneous nonlinear response is assumed as $P^{\mathrm{(NL)}}(t) = \chi^{(3)} E(t)^3$ with $\chi^{(3)} = 3.5$\,au (that is $1.66 \times 10^{-22}$\,m$^2$V$^{-2}$ in SI units \cite{RN249}). The UPPE reads
\begin{equation}
\partial_z \hat{E} = 
i \left( \frac{\omega}{u} -K \right)\hat{E}  
- \frac{2 \pi \omega}{K c^2} \left(i \omega \hat{P}^{\mathrm{(NL)}} \right),
\end{equation}
with $K = \sqrt{n_R^2 \frac{\omega^2}{c^2} - k_x^2}$. Numerical tables are used for the refractive index $n_R$, $c$ is the speed of light and $u$ is the group velocity of the NIR pulse. This includes the linear optical response via the refractive index $n_R$ of the substrate. The hat symbol indicates the Fourier transform in the dimensions of time and transverse space. In addition to the propagation direction $z$, one transverse dimension (the $x$-dimension) is included to account for the noncollinear geometry.  The electric field is treated as scalar field, because all pulses are polarized perpendicular to the plane of incidence. 

For the Ir-film, both the linear and the nonlinear optical response are contained in the expressions for $P$ and $J$ in (\ref{equ:P}) and (\ref{equ:J}) and the UPPE is
\begin{equation}
\partial_z \hat{E} = 
i \left( \frac{\omega}{u} -K \right)\hat{E}  
- \frac{2 \pi \omega}{K c^2} \left(i \omega \hat{P} + \hat{J} \right),
\end{equation}
with $K = \sqrt{\frac{\omega^2}{c^2} - k_x^2}$.

The UPPE is integrated numerically using the split-step method with an $x$-grid with 81 points in the interval [-250, 250]\,$\mu$m and a $z$-grid with 41 points in the interval [0, 80]\,$\mu$m for the propagation through the substrate and
8 points in the interval [0, 5]\,nm for the propagation through the Ir-film. 

Subsequent to the propagation inside the sample, the light propagating collinearly to pulses U and V is calculated by
\begin{equation}
E(\omega) = \hat{E}(\omega, k_x)
\end{equation}
with $\tan(-\frac{\alpha}{2}) = \frac{k_x}{\sqrt{(\omega / c)^2 + (k_x)^2}}$. 

\subsection{Atomic layer deposition}

The ultrathin Ir films are prepared on half of the surface area of fused silica substrates by means of atomic layer deposition (ALD) technique using a SunALE R-200 ALD reactor (Picosun Oy, Masala, Finland). Iridium(III) acetylacetonate (Ir(acac)$_3$) and moleculer oxygen (O$_2$) as precursors. The deposition temperature was kept at 380°C. The ALD process parameters for growing Ir are as follows: 6 sec Ir(acac)$_3$ pulse/60 sec N$_2$ purge/2 sec O$_2$ pulse/6 sec N$_2$ purge, resulting in a growth per cycle (GPC) of 0.6 Å/cycle \cite{RN285, RN278}.

\section*{Acknowledgments} This project was supported primarily by the Deutsche Forschungsgemeinschaft (DFG, German Research Foundation) - project ID 398816777 -
via projects A1, B1, B3 in the Collaborative Research Centre 1375 "Nonlinear optics down to atomic scales" (NOA). DK is funded via DFG Priority Programme 1840 - project ID 281272215 - "Quantum Dynamics in Tailored Intense Fields" (QUTIF). PP and AS acknowledge funding by the Fraunhofer Society (FhG, Attract 066-601020).

\bibliography{Bloch_Bib}

\end{document}